\documentclass[aps,prl,groupedaddress,twocolumn,10pt,showpacs]{revtex4}
\usepackage[pdftex]{graphicx}
\usepackage{amsmath,amssymb,color,subfigure,yllmath}
\providecommand{\myheading}[1]{\textbf{#1}}

\begin{document}
\title{Divergence of Dynamical Conductivity at Certain Percolative Superconductor-Insulator Transitions}
\author{Yen Lee Loh}
\author{Rajesh Dhakal}
\author{John F. Neis}
\author{Evan M. Moen}
\affiliation{Department of Physics and Astrophysics, University of North Dakota, Grand Forks, ND  43202, USA}
\date{This version started 2012-6-20; touched 2013-6-26; compiled \today}
\begin{abstract}
Random inductor-capacitor (LC) networks can exhibit percolative superconductor-insulator transitions (SITs).
We use a simple and efficient algorithm to compute the dynamical conductivity $\sigma(\omega,p)$ of one type of LC network on large ($4000 \times 4000$) square lattices, where $\delta=p-p_c$ is the tuning parameter for the SIT.
We confirm that the conductivity obeys a scaling form, so that the characteristic frequency scales as $\Omega \propto \abs{\delta}^{\nu z}$ with $\nu z \approx 1.91$, the superfluid stiffness scales as $\Upsilon \propto \abs{\delta}^t$ with $t \approx 1.3$, and the electric susceptibility scales as $\chi_E \propto \abs{\delta}^{-s}$ with $s = 2\nu z - t \approx 2.52$.
In the insulating state, the low-frequency dissipative conductivity is exponentially small,
whereas in the superconductor, it is linear in frequency.
The sign of $\Im\sigma(\omega)$ at small $\omega$ changes across the SIT.
Most importantly, we find that right at the SIT 
	$\Re\sigma(\omega) \propto \omega^{t/\nu z-1} \propto \omega^{-0.32}$,
	so that the conductivity \emph{diverges} in the DC limit,
	in contrast with most other classical and quantum models of SITs.
\end{abstract}
\pacs{74.62.En,74.78.-w,74.81.-g,78.67.-}
%74.40.Kb	Quantum critical phenomena
%74.62.En	Effects of disorder
%74.78.-w	Superconducting films and low-dimensional structures
%74.81.-g	Inhomogeneous superconductors and superconducting systems, including electronic inhomogeneities
%74.81.Bd	Granular, melt-textured, amorphous, and composite superconductors
%74.81.Fa	Josephson junction arrays and wire networks (see also 85.25.Cp Josephson devices)
%78.	Optical properties, condensed-matter spectroscopy and other interactions of radiation and particles with condensed matter
%78.67.-n	Optical properties of low-dimensional, mesoscopic, and nanoscale materials and structures (for magnetic properties of nanostructures, see 75.75.-c; for electronic transport in nanoscale structures, see 73.63.-b; for mechanical properties of nanoscale systems, see 62.25.-g)
%78.67.Pt	Multilayers; superlattices; photonic structures; metamaterials (see also 81.05.Xj, Metamaterials for chiral, bianisotropic and other complex media)
%78.67.Sc	Nanoaggregates; nanocomposites
\maketitle

Materials are classified as superconductors, metals, or insulators based on the way they respond to an electric field.
A thin film of a superconducting material can be turned into an insulator by increasing the disorder, decreasing the thickness\cite{haviland1989}, applying a parallel\cite{adams2004} or perpendicular magnetic field\cite{hebard1990}, or changing the gate voltage\cite{bollinger2011,lee2011}.
As a quantum phase transition occurring at zero temperature, this superconductor-insulator transition (SIT) has attracted much interest.
Early work focused on the most easily measurable quantity, the DC conductivity.\cite{goldmanMarkovicPhysicsToday1998,gantmakher2010,haviland1989,hebard1990,shahar1992, adams2004,steiner2005,stewart2007}
Recently, due to the availability of local scanning probes, attention has turned to the tunneling behavior \cite{sacepe2008,sacepe2010,sacepe2011,mondal2011,bouadim2011}.
In the case of the disorder-tuned SIT\cite{bouadim2011}, it has become clear that the SIT is ultimately due to a bosonic mechanism \cite{fisher1990} rather than a fermionic one \cite{finkelstein1994}.
The last step towards a full characterization of the SIT is to develop an understanding of the behavior of the AC conductivity.  This is a powerful probe of fluctuations on both long and short length and time scales, and it is of great interest especially as recent technological developments begin to open up more windows of the electromagnetic spectrum for measurement\cite{hetel2007,liu2011,hofmann2010,valdesaguilar2010}.

	%&&&&&&&&&&&&&&&&&&&&&&&&&&&&&&&&&&&&&&&&&&&&&&&&&&&&&&&&&&&&&&&&&&&&&&&&&&&&
	% FIGURE LijCiReSigmaLogLogPlot
	%&&&&&&&&&&&&&&&&&&&&&&&&&&&&&&&&&&&&&&&&&&&&&&&&&&&&&&&&&&&&&&&&&&&&&&&&&&&&
	\begin{figure}[!h]
	\includegraphics[width=0.99\columnwidth]{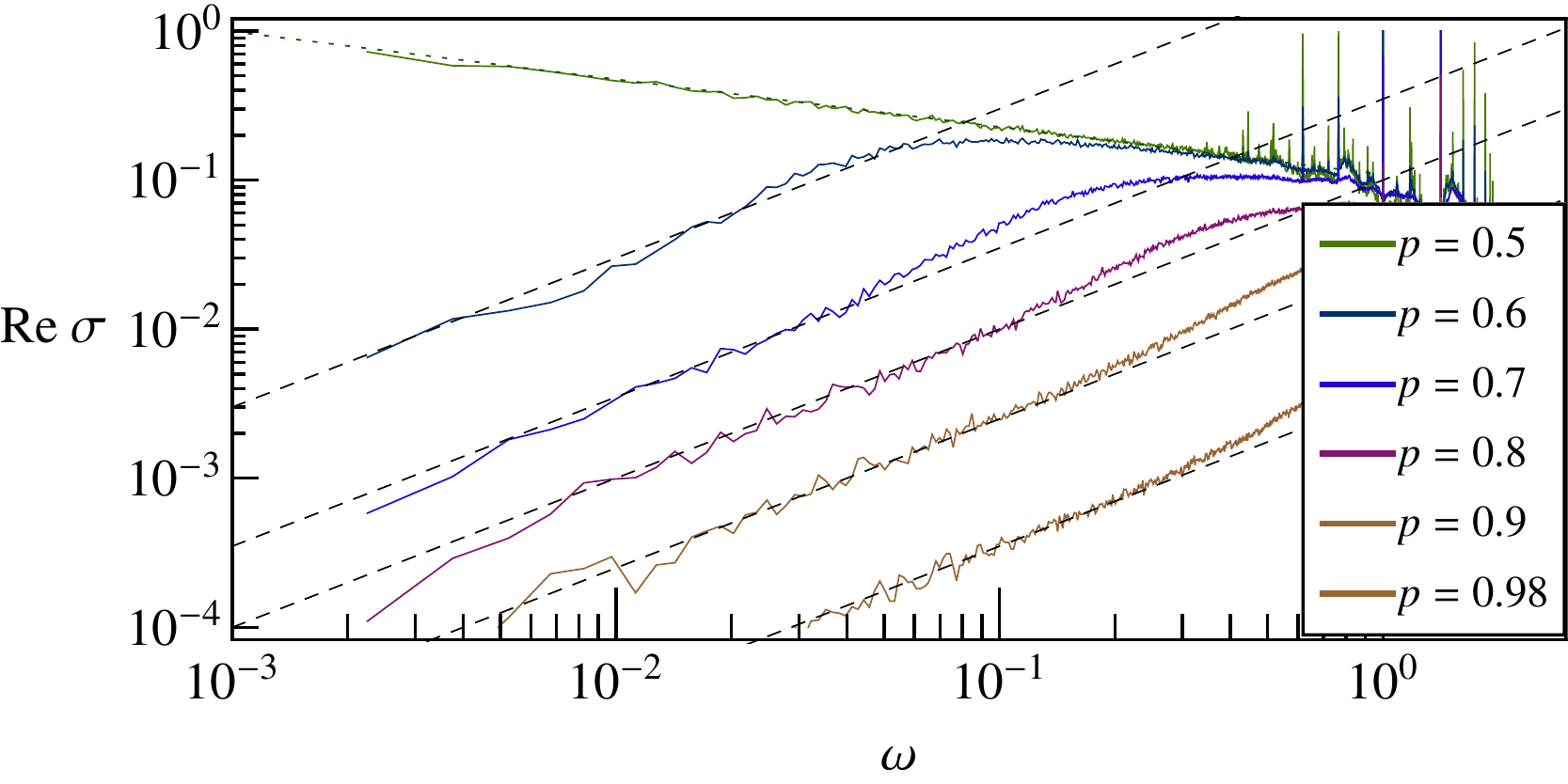}
	\includegraphics[width=0.99\columnwidth]{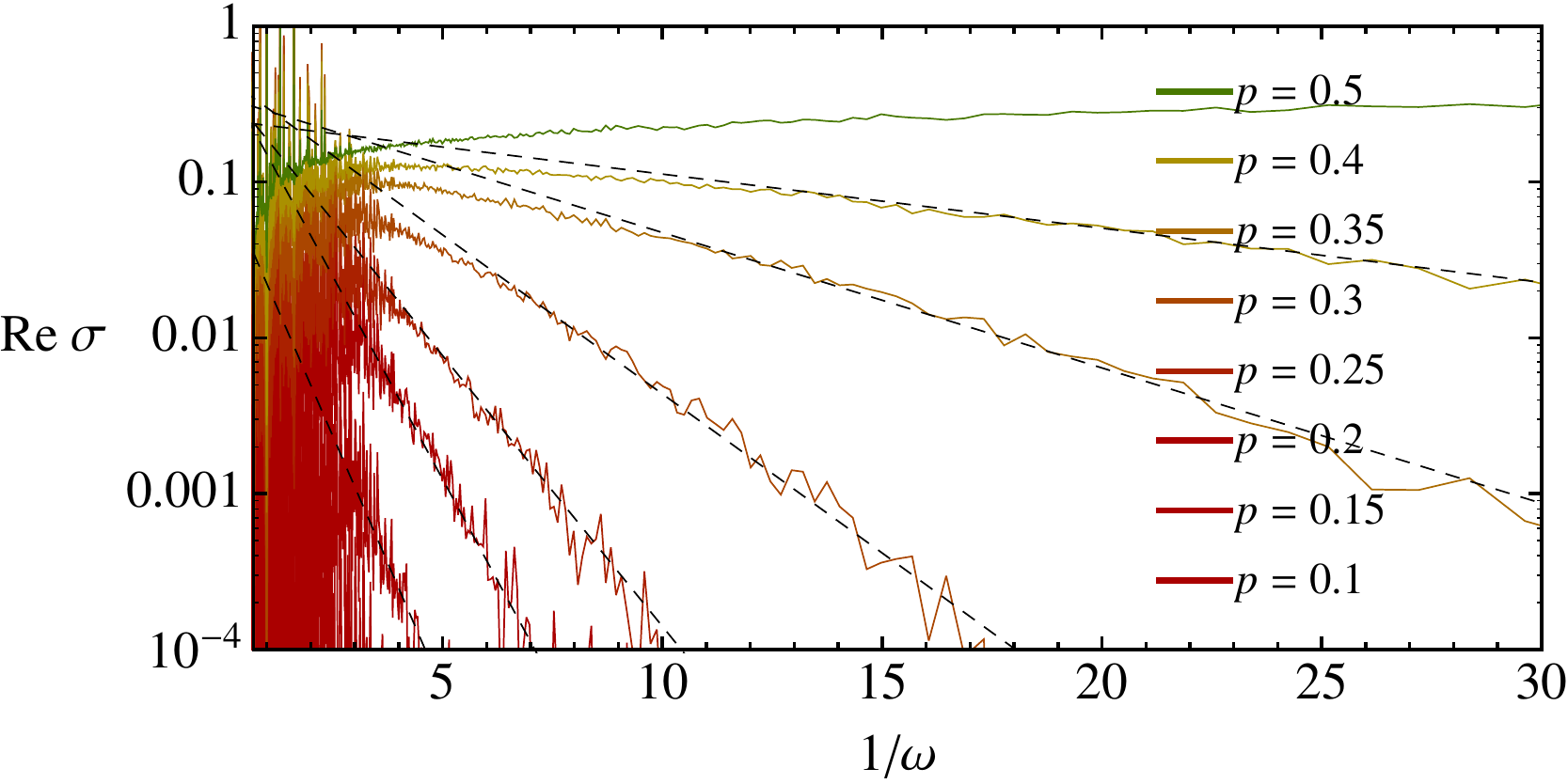}
	\\
	\caption{	
	Dissipative conductivity 
			$\Re\sigma(\omega)$ for the $L_{ij} C_{i}$ model.
	(Top)	
		Right at the SIT ($p=p_c=0.5$),
			$\Re\sigma$ diverges as $\omega^{-0.32}$ at small $\omega$,
			indicated by a downward-sloping straight line on the log-log plot.
		In the superconducting state ($p>p_c$),
			$\Re\sigma$ obeys an $\omega^1$ power law.
			The delta function $\delta(\omega)$ is not shown.
	(Bottom)
		In the insulating state ($p<p_c$), 
			$\Re\sigma$ is exponentially small ($e^{-\Omega(p)/\omega}$),
			such that a log plot of $\Re\sigma(\omega)$ vs $1/\omega$ shows
			straight lines.
	\label{LijCiReSigmaDivergence}
	}
	\end{figure}

	%&&&&&&&&&&&&&&&&&&&&&&&&&&&&&&&&&&&&&&&&&&&&&&&&&&&&&&&&&&&&&&&&&&&&&&&&&&&&
	% FIGURE ArrayPlots
	%&&&&&&&&&&&&&&&&&&&&&&&&&&&&&&&&&&&&&&&&&&&&&&&&&&&&&&&&&&&&&&&&&&&&&&&&&&&&
	\begin{figure*}[!thb]
	\includegraphics[width=0.95\columnwidth]{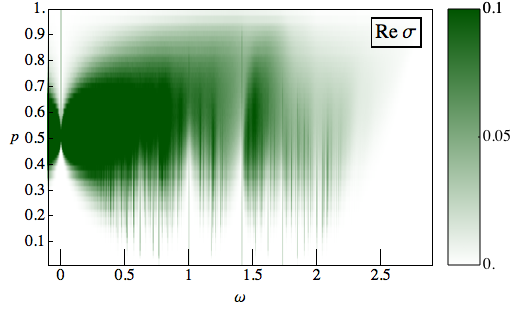}
	\includegraphics[width=0.95\columnwidth]{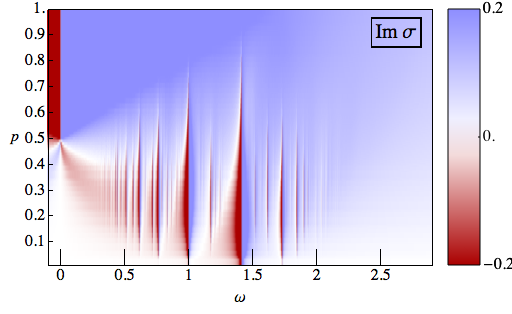}
	\caption{	
	Dynamical conductivity
		for the $L_{ij} C_{i}$ model
		for $p=0.02, 0.04, \dotsc, 0.40, 0.41, \dotsc, 0.60, 0.62, \dotsc, 0.98$
		for single realizations on a $4000\times 4000$ lattice.
	Red, green, and blue indicate
			insulating (capacitive),
			metallic (resistive),
			and superconducting (inductive) behavior respectively.
	The SIT occurs at the percolation threshold ($p_c=0.5$).
	The dissipative conductivity $\Re\sigma(\omega,p)$ 
		has a delta function in the superconductor ($p>p_c$),
		visible as a thin green line;
		most of the remaining weight occurs above a characteristic frequency 
		$\Omega \propto (p-p_c)^{\nu z}$,
		forming a fan shape reminiscent of quantum criticality.
	The reactive conductivity $\Im\sigma(\omega,p)$ changes sign	
		from capacitive to inductive
		as $p$ increases.
		\label{ArrayPlots}
	}
	\end{figure*}

One of the most important questions concerns the AC conductivity in the ``collisionless DC'' limit 
\cite{damle1997,sachdev_qpt}
\footnote{
In general one must be careful to distinguish between $\sigma^*$ and the true DC conductivity $\sigma^{DC} = \lim_{T\rightarrow 0}  \lim_{\omega\rightarrow 0} \sigma(\omega,T)$.
},
$\sigma^* = \lim_{\omega\rightarrow 0} \lim_{T\rightarrow 0} \sigma(\omega,T)$.
This has been the subject of a large body of work, including analytical arguments involving charge-vortex duality arguments, quantum Monte Carlo calculations in various representations, and experiments \cite{fisher1990,cha1991,sorensen1992,runge1992,makivic1993,batrouni1993,cha1994,smakov2005,linSorensen2011,sachdev_qpt,haviland1989,bollinger2011}.
It is often claimed that at the SIT $\sigma^*$ is finite and takes a universal value of the order of $\sigma_Q = 4e^2/h$, but there are large discrepancies between the ``universal'' values from various studies, and it is not clear whether there really is a universal value.

In this Letter we study the limit of a coarse-grained superconductor-insulator composite (e.g., millimeter-sized superconducting particles deposited on an insulating substrate).  In this situation quantum phase fluctuations are negligible and the SIT is governed by classical percolation.
We find that for one of the simplest inductor-capacitor network models, $\sigma(\omega)$ diverges as $\omega\rightarrow 0$, so that $\sigma^*$ is infinite!  See Fig.~\ref{LijCiReSigmaDivergence}.  
This is a surprising and important result, especially in the light of prior work on quantum models as well as classical models \cite{stroud1994}.
We also elucidate the structure of $\sigma(\omega)$ in this model, addressing the static limit, characteristic frequency, reactive response, and low-frequency contributions from rare regions in the insulator and from Goldstone modes in the superconductor.  Our predictions illustrate the range of behavior that can be obtained from classical percolation.  This providing a baseline that will very useful for comparing with quantum Monte Carlo results and with upcoming experiments, thus separating the effects of quenched disorder from those of quantum phase fluctuations.

%============================================================================
\myheading{Model:}
%============================================================================
In this Letter we focus on what we call the $L_{ij} C_{i}$ model.  This model contains capacitances-to-ground $C_i = C_0$ on every site $i$ and inductances $L_{ij} = L_0$ along bonds $ij$ with probability $p$.  Formally, the Lagrangian may be written as 
	\begin{align}
	\calL &= \sum_i \half C_i \dot{\theta}_i{}^2
	-	\sum_{\mean{ij}} \half L_{ij} {}^{-1}  (\theta_i - \theta_j + A_{ij})^2
	\label{Lagrangian}
	\end{align}
where $\theta_i$ are electromagnetic phase variables such that $V_i = \dot{\theta}_i$ and $A_{ij}$ is the external vector potential integrated along bond $ij$.  Note that this LC model is  almost the same as a Josephson junction array, except that it lacks Coulomb blockade effects resulting from charge quantization.  The dynamical electromagnetic response $\Upsilon=-dj/dA$, conductivity $\sigma=dj/dE$, and electric susceptibility $\chi_E=dP/dE$ can be defined in the usual way for a 2D system.

%============================================================================
\myheading{Methods:}
%============================================================================
Previous authors have attacked similar problems numerically using a matrix formalism\cite{jonckheere1998}, transfer matrix methods\cite{derrida1982,herrmann1984,zhangStroud1995}, and the Frank-Lobb bond propagation algorithm\cite{frank1988,zengStroud1989}.  In this study we employ a variant of the equation-of-motion method\cite{williams1985}, which is much simpler, more general, and more efficient.
Our approach is the theoretical analogue of Fourier transform nuclear magnetic resonance (FT-NMR) spectroscopy, where the frequency response is inferred from the free induction decay signal -- the impulse response in the time domain.
We apply a transient uniform electric field $E_x(t) = \delta(t)$, evolve currents and voltages according to the dynamical Kirchhoff equations, record the uniform component of the current $I_x(t)$, and extract the dissipative conductivity $\Re \sigma(\omega)$ using a fast Fourier transform.  The discretization error in the time evolution enters entirely in the form of systematic phase error, which we eliminate by a suitable transformation of the frequency variable.
This procedure can be shown to be formally equivalent to Chebyshev methods such as the kernel polynomial method \cite{wangChebyshev1994,silver1994,silver1996}.
The only source of error is the finite duration of the simulation, which leads to a finite frequency resolution.  We use a Kaiser window function that gives $\Re\sigma(\omega)$ with sidelobe amplitude below $\Delta \sigma \approx 10^{-8}$ and main lobe width $\Delta \omega \approx \frac{15\omega_\text{max}}{M}$ (where $M$ is the number of timesteps).  This prevents exponentially small tails in the spectrum from being contaminated by spectral leakage.
We compute $\Im\sigma(\omega)$ using a Kramers-Kronig transformation.  We estimate the superfluid stiffness $\Upsilon(\omega)$ from the weight in the lowest-frequency bin, and the electric susceptibility from
	$\chi_E = \int_0^{\infty} d\omega~ 2 \Re \sigma(\omega) / \omega^2$.
Detailed algorithms will be published elsewhere.

We simulated $4000\times 4000$ lattices for 40000 timesteps (i.e., 40000 Chebyshev moments), giving a resolution of $\Delta\omega \approx 0.0015$ after windowing and rebinning.
We quote angular frequency $\omega$ in units of $1/\sqrt{L_0C_0}$ and 2D conductivity (sheet conductance) $\sigma$ in units of $\sqrt{C_0/L_0}$.

	%&&&&&&&&&&&&&&&&&&&&&&&&&&&&&&&&&&&&&&&&&&&&&&&&&&&&&&&&&&&&&&&&&&&&&&&&&&&&
	% FIGURE LCStaticScaling
	%&&&&&&&&&&&&&&&&&&&&&&&&&&&&&&&&&&&&&&&&&&&&&&&&&&&&&&&&&&&&&&&&&&&&&&&&&&&&
	\begin{figure}
	\subfigure{
		\includegraphics[width=0.99\columnwidth]{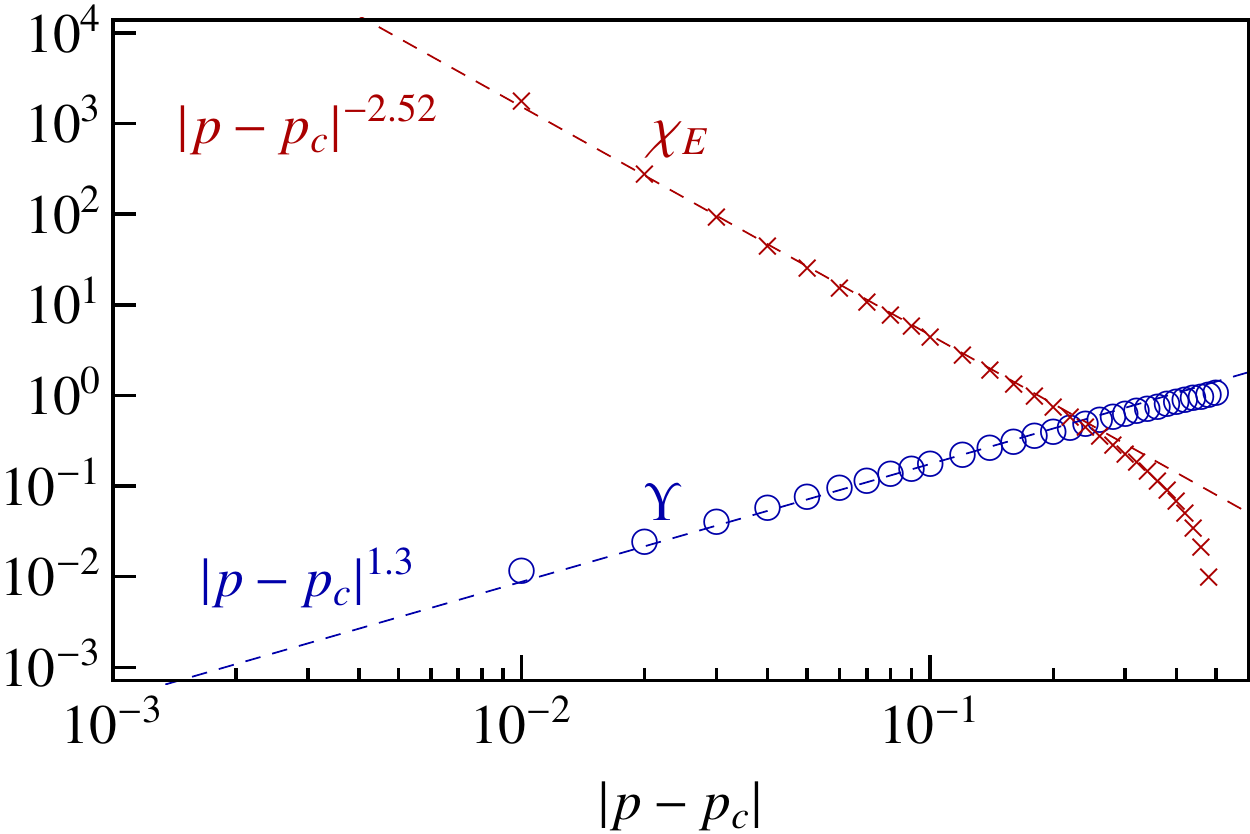}
	}
	\\
	\caption{	
		Static response functions of $LC$ networks 
			as a function of superconducting bond fraction $p$.
		The insulating state is characterized by a finite electric susceptibility $\chi_E$,
			which diverges at the percolation transition.
		The superconducting state is characterized by a finite superfluid stiffness $\Upsilon$,
			which vanishes at the transition.		
		Near the percolation threshold $p$,
			the superfluid stiffness scales as
			$\Upsilon \sim (p - p_c)^{1.30}$.
		For the $L_{ij} C_{ij}$ model near $p_c$,
			$\chi_E = \Upsilon^{-1} \sim (p_c - p)^{-1.30}$ due to duality.
		For the $L_{ij} C_{i}$ model near $p_c$, 
			$\chi_E \sim (p_c - p)^{-2.7}$.
			\label{LCStaticScaling}
	}
	\end{figure}

	%&&&&&&&&&&&&&&&&&&&&&&&&&&&&&&&&&&&&&&&&&&&&&&&&&&&&&&&&&&&&&&&&&&&&&&&&&&&&
	% FIGURE Collapse
	%&&&&&&&&&&&&&&&&&&&&&&&&&&&&&&&&&&&&&&&&&&&&&&&&&&&&&&&&&&&&&&&&&&&&&&&&&&&&
	\begin{figure*}
	\includegraphics[width=0.95\columnwidth]{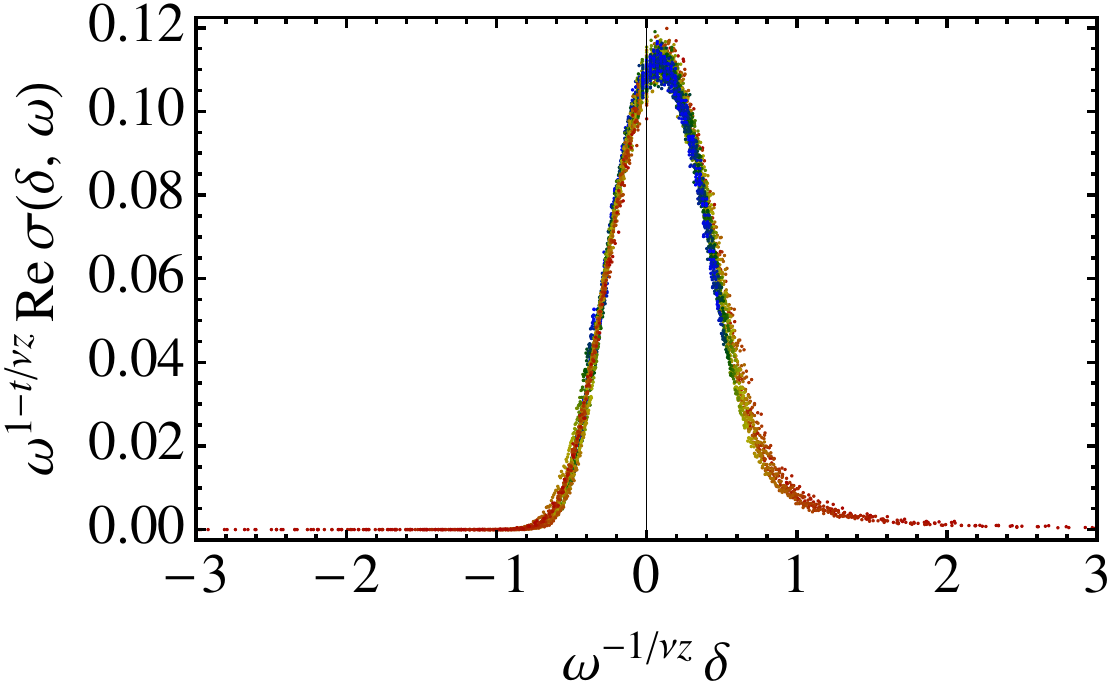}
	\includegraphics[width=0.95\columnwidth]{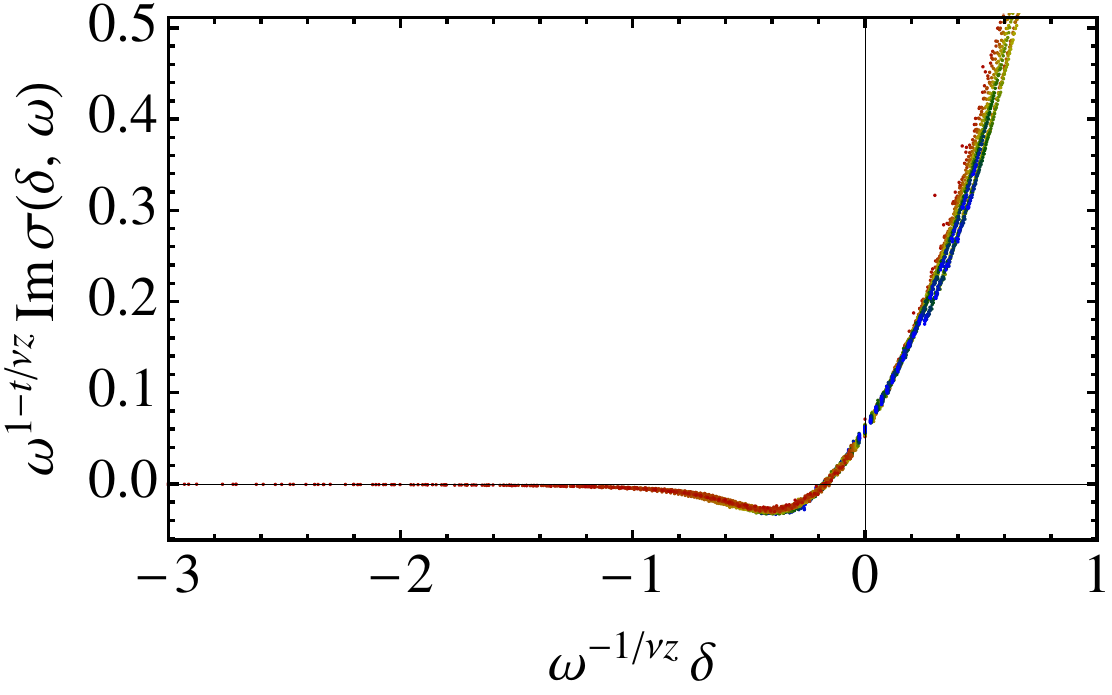}
	\caption{			
		Real and imaginary parts of the scaling function $f(x)$
			extracted using scaling collapse	 of $\sigma(p,\omega)$
			for $0.3\leq p \leq 0.7$ and $0 < \omega < 0.2$.
		This confirms the scaling form
			$
			\sigma(\omega,\delta) = \omega^{t/\nu z - 1} f( \omega^{-1/\nu z} \delta  )
			$
		where $\delta = p-p_c$, $t=1.30$, and $\nu z=1.91$.
		Colors of data points indicate $\omega$ values.
		We have verified that scaling collapse is good
			for $10^{-4} < f(x) < 10^1$,
			beyond the dynamic range shown above.
		\label{Collapse}
	}
	\end{figure*}

%============================================================================
\myheading{Results:}
%============================================================================
Color plots of $\Re\sigma(\omega,p)$ and $\Im\sigma(\omega,p)$ are shown in Fig.~\ref{ArrayPlots}. The data contain a wealth of interesting information that we list below.

%----------------------------------------------------------------------------
\myheading{(a) Superfluid stiffness:}
%----------------------------------------------------------------------------
On the superconducting side of the SIT ($p>p_c$), the superfluid stiffness scales as $\Upsilon(p) \sim \delta^t$ where $\delta=p-p_c$ and $t\approx 1.30$, as shown in Fig.~\ref{LCStaticScaling}.  This agrees with results for resistor networks \cite{lobb1984,hong1984}.

%----------------------------------------------------------------------------
\myheading{(b) Characteristic frequency:}
%----------------------------------------------------------------------------
For 2D percolation the correlation length diverges as $\xi \sim \abs{\delta}^{-\nu}$ with $\nu=4/3$ \cite{denNijs1979}.  This suggests that there is a characteristic (angular) frequency $\Omega \sim \xi^{-z} \sim \abs{\delta}^{\nu z}$, where $z$ is a dynamical critical exponent.  Indeed, Fig.~\ref{ArrayPlots} suggests that most of the spectral weight in $\Re\sigma(\omega)$ occurs above a frequency $\Omega \sim \delta^{\nu z}$, forming a shape analogous to a ``quantum critical'' fan.

%----------------------------------------------------------------------------
\myheading{(c) Divergent conductivity at SIT:}
%----------------------------------------------------------------------------
At the SIT ($p=p_c$), $\Re\sigma(\omega)$ does not tend to a finite limit as in many other systems, but instead it diverges!  This is illustrated in Fig.~\ref{LijCiReSigmaDivergence}.  
We find a very good fit to a power law of the form $\Re\sigma(\omega) \approx \omega^{-a}$ where $a\approx 0.32(1)$.

%----------------------------------------------------------------------------
\myheading{(d) Scaling collapse:}
%----------------------------------------------------------------------------
Based on observations (a) and (b), we postulate the scaling form
$
\sigma(\omega,\delta) = \omega^{t/\nu z - 1} f( \omega^{-1/\nu z} \delta  )
$.
This mandates that 
$\sigma(\omega) \propto \omega^{t/\nu z-1}$ at the SIT.
Comparing with (c), we see that we must have 
$t/\nu z - 1 = -a$, so that 
$\nu z = t/(1-a) \approx 1.91$.
Indeed, we find that both $\Re\sigma$ and $\Im\sigma$ collapse onto single curves for $\nu z\approx 1.91$, as shown in Fig.~\ref{Collapse}.  The details of $f(x)$ will be reported elsewhere.

%----------------------------------------------------------------------------
\myheading{(e) Electric susceptibility:}
%----------------------------------------------------------------------------
On the insulating side of the SIT ($p<p_c$), the scaling form dictates that the electric susceptibility must scale as $\chi_E \sim \abs{\delta}^{-s}$ with $s = 2\nu z - t \approx 2.52$.  Indeed, Fig.~\ref{LCStaticScaling} shows a good fit to this power law.

%----------------------------------------------------------------------------
\myheading{(f) Low-frequency dissipation:}
%----------------------------------------------------------------------------
The characteristic frequency $\Omega$ does \emph{not} correspond to a hard gap in the spectrum.  In the insulating state, large rare regions contribute exponentially small weight to $\Re\sigma$ all the way down to zero frequency, as shown in the bottom panel of Fig.~\ref{LijCiReSigmaDivergence}.  This is a ramification of Griffiths-McCoy-Wu physics\cite{mccoy1968,griffiths1969,vojta2010} in systems with quenched disorder.
The superconducting state has linear low-frequency dissipation $\Re\sigma \sim \omega$, as shown in the top panel of Fig.~\ref{LijCiReSigmaDivergence}.  We believe this is due to the excitation of acoustic ``transmission-line'' modes that is permitted in the presence of disorder.  (It is also reminiscent of Ref.~\onlinecite{pollak1972}.)

%----------------------------------------------------------------------------
\myheading{(g) Reactive conductivity:}
%----------------------------------------------------------------------------
The reactive part of the dynamical conductivity, $\Im \sigma(\omega)$, is shown in Fig.~\ref{ArrayPlots} and implicitly in Fig.~\ref{Collapse}.  
In the insulating state, the low-frequency response is capacitive ($\Im\sigma(\omega) < 0$).
In the superconducting state, it is inductive ($\Im\sigma(\omega) > 0$).  
This suggests that the sign of $\Im \sigma(\omega)$ may be used in experiments as a criterion to distinguish between insulating and superconducting states.

%----------------------------------------------------------------------------
\myheading{Importance of on-site capacitances:}
%----------------------------------------------------------------------------
In previous studies of the dynamical conductivity of classical systems near percolation, the insulating (capacitive) elements were placed along bonds, in series with the superconducting (inductive) elements \cite{stroud1994,jonckheere1998}.  We have studied an $LC$ model of this type, which we call the $L_{ij} C_{ij}$ model.  We find that for that model $\nu z=t \approx 1.3$, so that $\sigma^*$ is finite and $\chi_E$ scales differently.  In addition, $\Re\sigma$ has exponentially small low-frequency dissipation in the superconducting state, and there are various peculiarities due to self-duality \cite{straley1976}.  Details will be published elsewhere.
Ultimately, we feel that the $L_{ij} C_{i}$ model is likely to be more generic than the $L_{ij} C_{ij}$ model for various reasons.  Furthermore, as mentioned earlier, the $L_{ij} C_{i}$ model is the limiting case of a Josephson junction array when the charging energy is negligible.

%============================================================================
\myheading{Conclusions:}
%============================================================================
We have studied the percolative superconductor-insulator transition in two-dimensional classical LC networks, in particular, the $L_{ij} C_{i}$ model.
We used an efficient algorithm to compute $\sigma(\omega,p)$ on large lattices ($4000 \times 4000$ sites).
We find the critical exponents $t \approx 1.30$ (in agreement with results on resistor networks), $\nu z \approx 1.91$, and $s = 2\nu z - t \approx 2.52$.
We have extracted the complex-valued scaling function.
In the insulating state, the low-frequency dissipative conductivity is exponentially small,
whereas in the superconductor, it is linear in frequency.
The sign of $\Im\sigma(\omega)$ at small $\omega$ changes across the SIT.
Most surprisingly, right at the SIT, $\Re\sigma(\omega)$ diverges as $\omega\rightarrow 0$.

As remarked in the introduction, most studies on quantum models and classical models report finite values of $\sigma^*$ at the SIT.   In this light, it is extremely surprising that $\sigma^*$ is divergent for a simple classical model!  
Our results form an important baseline to which to compare simulations of more complicated models such as XY, Bose-Hubbard, and Fermi-Hubbard models, thus allowing one to separate the effects of quenched disorder, quantum phase fluctuations (Coulomb blockade physics), and pairbreaking physics.

We gratefully acknowledge Mason Swanson and Mohit Randeria for useful discussions.
%We gratefully acknowledge support from nobody.

%============================================================================
% Bibliography
%============================================================================

%==== If using BibTeX ====
%\bibliography{sit}

\begin{thebibliography}{10}
\newcommand{\enquote}[1]{``#1''}

\bibitem{haviland1989}
D.~B. Haviland, Y.~Liu, and A.~M. Goldman, PRL {\bf 62}, 2180 (1989).

\bibitem{adams2004}
P.~Adams, Phys. Rev. Lett. {\bf 92}, 067003 (2004).

\bibitem{hebard1990}
A.~F. Hebard and M.~A. Paalanen, Phys. Rev. Lett. {\bf 65}, 927 (1990).

\bibitem{bollinger2011}
A.~T. Bollinger, G.~Dubuis, J.~Yoon, D.~Pavuna, J.~Misewich, and I.~Bozovic,
  Nature {\bf 472}, 458 (2011), ISSN 0028-0836.

\bibitem{lee2011}
Y.~Lee, C.~Clement, J.~Hellerstedt, J.~Kinney, L.~Kinnischtzke, X.~Leng, S.~D.
  Snyder, and A.~M. Goldman, Phys. Rev. Lett. {\bf 106}, 136809 (2011).

\bibitem{goldmanMarkovicPhysicsToday1998}
A.~Goldman and N.~Markovic, Physics Today {\bf 51}, 39 (1998).

\bibitem{gantmakher2010}
V.~F. Gantmakher and V.~T. Dolgopolov, Physics-Uspekhi {\bf 53}, 3 (2010).

\bibitem{shahar1992}
D.~Shahar and Z.~Ovadyahu, Phys. Rev. B {\bf 46}, 10917 (1992).

\bibitem{steiner2005}
M.~A. Steiner, G.~Boebinger, and A.~Kapitulnik, Phys. Rev. Lett. {\bf 94},
  107008 (2005).

\bibitem{stewart2007}
M.~D. Stewart, A.~Yin, J.~M. Xu, and J.~M. Valles, Science {\bf 318}, 1273
  (2007).

\bibitem{sacepe2008}
B.~Sac{\'e}p{\'e}, C.~Chapelier, T.~I. Baturina, V.~M. Vinokur, M.~R. Baklanov,
  and M.~Sanquer, PRL {\bf 101}, 157006 (2008).

\bibitem{sacepe2010}
B.~Sac{\'e}p{\'e}, C.~Chapelier, T.~I. Baturina, V.~M. Vinokur, M.~R. Baklanov,
  and M.~Sanquer, Nature Communications {\bf 1}, 140 (2010).

\bibitem{sacepe2011}
B.~Sac{\'e}p{\'e}, T.~Dubouchet, C.~Chapelier, M.~Sanquer, M.~Ovadia,
  D.~Shahar, M.~Feigel'man, and L.~Ioffe, Nat. Phys. {\bf 7}, 239 (2011), ISSN
  1745-2473.

\bibitem{mondal2011}
M.~Mondal, A.~Kamlapure, M.~Chand, G.~Saraswat, S.~Kumar, J.~Jesudasan,
  L.~Benfatto, V.~Tripathi, and P.~Raychaudhuri, Phys. Rev. Lett. {\bf 106},
  047001 (2011).

\bibitem{bouadim2011}
K.~Bouadim, Y.~L. Loh, M.~Randeria, and N.~Trivedi, Nat. Phys. {\bf 7}, 884
  (2011).

\bibitem{fisher1990}
M.~P.~A. Fisher, G.~Grinstein, and S.~M. Girvin, PRL {\bf 64}, 587 (1990).

\bibitem{finkelstein1994}
A.~M. Finkel'stein, Physica B {\bf 197}, 636 (1994).

\bibitem{hetel2007}
I.~Hetel, T.~R. Lemberger, and M.~Randeria, Nat. Phys. {\bf 3}, 700 (2007),
  ISSN 1745-2473.

\bibitem{liu2011}
W.~Liu, M.~Kim, G.~Sambandamurthy, and N.~P. Armitage, Phys. Rev. B {\bf 84},
  024511 (2011).

\bibitem{hofmann2010}
T.~Hofmann, C.~M. Herzinger, A.~Boosalis, T.~E. Tiwald, J.~A. Woollam, and
  M.~Schubert, Review of Scientific Instruments {\bf 81}, 023101 (2010), ISSN
  00346748.

\bibitem{valdesaguilar2010}
R.~Vald\'es~Aguilar, L.~S. Bilbro, S.~Lee, C.~W. Bark, J.~Jiang, J.~D. Weiss,
  E.~E. Hellstrom, D.~C. Larbalestier, C.~B. Eom, and N.~P. Armitage, Phys.
  Rev. B {\bf 82}, 180514 (2010).

\bibitem{damle1997}
K.~Damle and S.~Sachdev, Phys. Rev. B {\bf 56}, 8714 (1997).

\bibitem{sachdev_qpt}
S.~Sachdev, {\em Quantum Phase Transitions\/} (Cambridge, London, 1999).

\bibitem{Note1}
In general one must be careful to distinguish between $\sigma ^*$ and the true
  DC conductivity $\sigma ^{DC} = \protect \qopname \relax m{lim}_{T\rightarrow
  0} \protect \qopname \relax m{lim}_{\omega \rightarrow 0} \sigma (\omega
  ,T)$.

\bibitem{cha1991}
M.-C. Cha, M.~P.~A. Fisher, S.~M. Girvin, M.~Wallin, and A.~P. Young, Phys.
  Rev. B {\bf 44}, 6883 (1991).

\bibitem{sorensen1992}
E.~S. S\o{}rensen, M.~Wallin, S.~M. Girvin, and A.~P. Young, Phys. Rev. Lett.
  {\bf 69}, 828 (1992).

\bibitem{runge1992}
K.~J. Runge, Phys. Rev. B {\bf 45}, 13136 (1992).

\bibitem{makivic1993}
M.~Makivi\ifmmode~\acute{c}\else \'{c}\fi{}, N.~Trivedi, and S.~Ullah, Phys.
  Rev. Lett. {\bf 71}, 2307 (1993).

\bibitem{batrouni1993}
G.~G. Batrouni, B.~Larson, R.~T. Scalettar, J.~Tobochnik, and J.~Wang, Phys.
  Rev. B {\bf 48}, 9628 (1993).

\bibitem{cha1994}
M.-C. Cha and S.~M. Girvin, Phys. Rev. B {\bf 49}, 9794 (1994).

\bibitem{smakov2005}
J.~\ifmmode~\check{S}\else \v{S}\fi{}makov and E.~S\o{}rensen, Phys. Rev. Lett.
  {\bf 95}, 180603 (2005).

\bibitem{linSorensen2011}
F.~Lin, E.~S. S\o{}rensen, and D.~M. Ceperley, Phys. Rev. B {\bf 84}, 094507
  (2011).

\bibitem{stroud1994}
D.~Stroud, Physica A: Statistical Mechanics and its Applications {\bf 207}, 280
  (1994), ISSN 0378-4371.

\bibitem{jonckheere1998}
T.~Jonckheere and J.~M. Luck, Journal of Physics A: Mathematical and General
  {\bf 31}, 3687 (1998).

\bibitem{derrida1982}
B.~Derrida and J.~Vannimenus, Journal of Physics A: Mathematical and General
  {\bf 15}, L557 (1982).

\bibitem{herrmann1984}
H.~J. Herrmann, B.~Derrida, and J.~Vannimenus, Phys. Rev. B {\bf 30}, 4080
  (1984).

\bibitem{zhangStroud1995}
X.~Zhang and D.~Stroud, Phys. Rev. B {\bf 52}, 2131 (1995).

\bibitem{frank1988}
D.~J. Frank and C.~J. Lobb, Phys. Rev. B {\bf 37}, 302 (1988).

\bibitem{zengStroud1989}
X.~C. Zeng, P.~M. Hui, and D.~Stroud, Phys. Rev. B {\bf 39}, 1063 (1989).

\bibitem{williams1985}
M.~L. Williams and H.~J. Maris, Phys. Rev. B {\bf 31}, 4508 (1985).

\bibitem{wangChebyshev1994}
L.-W. Wang, Phys. Rev. B {\bf 49}, 10154 (1994).

\bibitem{silver1994}
R.~Silver and H.~R{\"o}der, International Journal of Modern Physics C {\bf 05},
  735 (1994).

\bibitem{silver1996}
R.~Silver, H.~Roeder, A.~Voter, and J.~Kress, Journal of Computational Physics
  {\bf 124}, 115  (1996), ISSN 0021-9991.

\bibitem{lobb1984}
C.~J. Lobb and D.~J. Frank, Phys. Rev. B {\bf 30}, 4090 (1984).

\bibitem{hong1984}
D.~C. Hong, S.~Havlin, H.~J. Herrmann, and H.~E. Stanley, Phys. Rev. B {\bf
  30}, 4083 (1984).

\bibitem{denNijs1979}
M.~P.~M. den Nijs, Journal of Physics A: Mathematical and General {\bf 12},
  1857 (1979).

\bibitem{mccoy1968}
B.~M. McCoy and T.~T. Wu, Phys. Rev. {\bf 176}, 631 (1968).

\bibitem{griffiths1969}
R.~B. Griffiths, Phys. Rev. Lett. {\bf 23}, 17 (1969).

\bibitem{vojta2010}
T.~Vojta, Journal of Low Temperature Physics {\bf 161}, 299 (2010), ISSN
  0022-2291, 10.1007/s10909-010-0205-4.

\bibitem{pollak1972}
M.~Pollak and G.~E. Pike, Physical Review Letters {\bf 28}, 1449 (1972).

\bibitem{straley1976}
J.~P. Straley, Journal of Physics C: Solid State Physics {\bf 9}, 783 (1976).

\end{thebibliography}

%==== If not using BibTeX ====

\end{document}